\documentclass[twocolumn,showpacs,amsmath,amssymb,10pt,aps]{revtex4}
\usepackage{graphicx,color}
\usepackage{amsmath}
\usepackage{amssymb}
\usepackage{bbm}
\usepackage{color}

\usepackage[utf8]{inputenc}

\usepackage{amsfonts}
\usepackage{bm}
\newcommand{\rr}{\mathbb{R}}

\newcommand{\ud}{\mathrm{d}}
\newcommand{\ot}{{\,\otimes\,}}
\newcommand{{\Cd}}{{\mathbb{C}^d}}

\def\oper{{\mathchoice{\rm 1\mskip-4mu l}{\rm 1\mskip-4mu l}
{\rm 1\mskip-4.5mu l}{\rm 1\mskip-5mu l}}}
\def\<{\langle}
\def\>{\rangle}
\newtheorem{Theorem}{Theorem}
\newtheorem{Lemma}{Lemma}

\newtheorem{property}{Property}
\newtheorem{Corollary}{Corollary}
\newtheorem{Definition}{Definition}
\newtheorem{Remark}{Remark}
\newtheorem{Example}{Example}
\newtheorem{Proposition}{Proposition}

\newcommand{\beq}{\begin{equation}}
\newcommand{\eeq}{\end{equation}}
\newcommand{\bear}{\begin{eqnarray}}
\newcommand{\ear}{\end{eqnarray}}
\newcommand{\bdm}{\begin{displaymath}}
\newcommand{\edm}{\end{displaymath}}

\newcommand{\lhs}{\mathrm{LHS}}
\newcommand{\rhs}{\mathrm{RHS}}

\begin{document}
\title{\textbf{Sufficient conditions for memory kernel master equation}}
\author{Dariusz Chru\'sci\'nski and Andrzej Kossakowski}
\affiliation{ Institute of Physics, Faculty of Physics, Astronomy and Informatics \\  Nicolaus Copernicus University,
Grudzi\c{a}dzka 5/7, 87--100 Toru\'n, Poland}



\begin{abstract}
We derive sufficient conditions for the memory kernel governing non-local master equation which guarantee legitimate (completely positive and trace-preserving) dynamical map. It turns out that these conditions provide a natural parameterizations of the dynamical map being a generalization  of Markovian semigroup. {This parametrization is defined by the so-called legitimate pair -- monotonic quantum operation and completely positive map -- and it is shown that such class of maps} cover almost all known examples from Markovian semigroup, semi-Markov evolution up to collision models and their generalization.
\end{abstract}

\maketitle


{\em Introduction.} --- The dynamics of open quantum systems attracts nowadays considerable attention \cite{Breuer,Weiss,Rivas-Huelga}.
It is relevant for a proper description of quantum evolution of the system in question coupled to an environment. Since any realistic system is never perfectly isolated it must be treated as an open one and hence the theory of open quantum systems plays the fundamental role for analyzing, modelling and controlling realistic quantum systems. It is, therefore, clear that open quantum systems are also fundamental for potential applications in modern quantum technologies
such as quantum communication, cryptography and computation \cite{QIT}.

There are two basic approaches to the evolution of open quantum systems. Time-local approach  based on the following master equation
\begin{equation}\label{ME}
  \dot{\rho}(t) = \mathcal{L}(t)\rho(t) ,
\end{equation}
where $\mathcal{L}(t)$ denotes time-local generator. This approach provides a straightforward generalization of the celebrated GKSL master equation \cite{GKS,L} derived exactly 40 years ago. It was proved \cite{GKS,L} that in order to generate legitimate (completely positive (CP) and trace-preserving (CPTP)) dynamical map $\rho \rightarrow \rho(t) = \Lambda(t)[\rho]$ the corresponding generator has the following canonical form
\begin{equation}\label{Lindblad}
  \mathcal{L}[\rho] = -i[H,\rho] + \sum_\alpha \gamma_\alpha (V_\alpha \rho V_\alpha^\dagger - \frac 12 \{ V_\alpha^\dagger V_\alpha,\rho\}) ,
\end{equation}
where $H$ denotes an effective Hamiltonian, $V_\alpha$ are noise operators, and $\gamma_\alpha \geq 0$ describe decoherence/dissipation rates. In the time-dependent case one replaces $H$, $V_\alpha$ and $\gamma_\alpha$ by the corresponding time-dependent objects. This approach is well suited to analyze the problem of Markovianity \cite{BLP,RIVAS,BLPV}. In particular, if $\gamma_\alpha(t) \geq 0$, then the map $\Lambda(t)$ is CP-divisible, i.e. $\Lambda(t) = V(t,s)\Lambda(s)$ and $V(t,s)$ is a collection of CPTP propagators for $t \geq s$ \cite{D1,D2,D3}.

The second approach is based on the non-local Nakajima-Zwanzig \cite{NZ} (see also \cite{NZ-inni,Haake})
\begin{equation}\label{NZ}
  \dot{\rho}(t) = \int_0^t K_{\rm NZ}(t-\tau) \rho(\tau) d\tau ,
\end{equation}
in which quantum memory effects are taken into account
through the introduction of the memory kernel $K_{\rm NZ}(t)$. This means that the rate of change of the state $\rho(t)$ at time $t$ depends
on its history (starting at t = 0).

The notorious problem faced by the non-local master equation (\ref{NZ}) is complete positivity of the evolution described by the map $\Lambda(t)$.  This problem was already observed by Barnett and Stenholm \cite{B-S} for the memory kernel $K(t)= k(t)\mathcal{L}$ with $k(t)$ being some memory function  and the legitimate Markovian generator $\mathcal{L}$. Lidar and Shabani \cite{Lidar} derived so called {\em post-Markovian master equation} which corresponds to
 $ K_{\rm LS}(t) = k(t) \mathcal{L} e^{ \mathcal{L} t}$
and shown \cite{Lidar} that it interpolates between the generalized measurement interpretation
of the exact Kraus operator sum map and the continuous
measurement interpretation of the Markovian dynamics. In the qubit case these proposals were analyzed in \cite{Maniscalco} and recently in \cite{Wudarski}. It should be stressed that both for the phenomenological kernel of Barnett and Stenholm and Lindar-Shabani post-Markovian kernel the complete positivity of the corresponding dynamical map is not guaranteed. It depends both on the memory function $k(t)$ and Markovian generator $\mathcal{L}$.
Other approaches were discussed e.g. in \cite{Budini,Wilke,AK,EPL}.

An interesting proposal for the quantum evolution governed by (\ref{NZ}) is provided by so-called semi-Markov evolution \cite{Budini,B-V,NJP-2011}. This proposal by construction leads to $\Lambda(t)$ which is always CPTP. Another approach -- the so-called {\em collision models} -- was advocated recently in \cite{Palma} (see also \cite{basano-1} for more general recent discussion and \cite{Campbell} for further analysis). 
%
Originally, collision models were proposed to derive Markovian master equation \cite{Col-1}. In this approach one obtains the standard Markovian semigroup dynamics without applying sophisticated Markovian approximations (like weak-coupling or singular coupling limit).
Such standard models may be called {\em  memoryless collision models}. Now, following \cite{Palma} one endows the bath (collection of identical ancillas) with memory by introducing inter-ancillary collisions between next system-ancilla interactions and finally one arrives .
at the following equation for the evolution of the system density operator $\rho(t)$
\begin{equation}\label{CM-ME}
  \dot{\rho}(t) = \Gamma \int_0^t d\tau e^{-\Gamma \tau} \mathcal{F}(\tau) \dot{\rho}(t-\tau) + e^{-\Gamma t} \dot{\mathcal{F}}(t)\rho_0\ ,
\end{equation}
where $\mathcal{F}(t)$ is a family of CPTP maps satisfying $\mathcal{F}(0) = \oper$, $\Gamma \geq 0$, and $\rho_0$ is a initial state of the system. Interestingly, this  integro-differential equation contains $\rho_0$ which is interpreted in \cite{Palma} as the strong signature of non-Markovian
behavior. As we show apparently different equations (\ref{NZ}) and (\ref{CM-ME}) are in fact deeply connected.

In this Letter we provide simple parametrization of legitimate memory kernels by a pair of CP maps $\{N(t),Q(t)\}$ satisfying one additional constraint (\ref{TR}) which is responsible for the trace-preservation of the corresponding dynamical map $\Lambda(t)$. A class of pairs enjoys a series of elegant properties: it is convex, consistent with reduction procedure and closed with respect to a natural set of local gauge and shift transformations. We show that it provides a generalization of Markovian semigroup and covers almost all known examples, e.g. semi-Markov evolution and so-called collision models.


\vspace{.2cm}

{\em A class of legitimate solutions.} --- Any memory kernel $\mathcal{K}_{\rm NZ}(t)$ has the following general structure
 $ K_{\rm NZ}(t) = B(t) - Z(t)$, 
where maps $B(t)$ and $Z(t)$ are Hermitian \cite{HER} and satisfy ${\rm Tr}(B_t[\rho]) =  {\rm Tr}(Z_t[\rho])$ which guarantees that $\mathcal{K}_{\rm NZ}(t)$ kills the trace and hence $\Lambda(t)$ is trace-preserving. It should be stressed  that any super-operator $K$ which annihilates the trace, i.e. ${\rm Tr}(K[\rho])=0$, has this form. Particular example is provided by Markovian generator (\ref{Lindblad}) where the splitting `$B-Z$' is {\em canonical}.
Passing to the Laplace Transform (LT) domain one transforms the Nakajima-Zwanzig equation for the dynamical map $\Lambda_t$ into the following equation
\begin{equation}\label{}
  \widetilde{\Lambda}(s) = \frac{1}{s - \widetilde{K}_{\rm NZ}(s)} = \frac{1}{s - \widetilde{B}(s) + \widetilde{Z}(s)} .
\end{equation}
Introducing $  \widetilde{N}(s) = [s + \widetilde{Z}(s)]^{-1}$,
one arrives at
 $ \widetilde{\Lambda}(s) = \widetilde{N}(s) + \widetilde{N}(s) \widetilde{B}(s) \widetilde{\Lambda}(s)$,
which after iteration leads to
\begin{equation}\label{series-s}
  \widetilde{\Lambda}(s) = \widetilde{N}(s) + \widetilde{N}(s) \widetilde{Q}(s) +  \widetilde{N}(s) \widetilde{Q}(s) \widetilde{Q}(s) + \ldots
\end{equation}
with $\widetilde{Q}(s) = \widetilde{B}(s) \widetilde{N}(s) $. Finally, assuming $||\widetilde{Q}(s)||_1 < 1$ \cite{norm} the above series converges to
\begin{equation}\label{!}
  \widetilde{\Lambda}(s) = \widetilde{N}(s)[1 - \widetilde{Q}(s)]^{-1} \ .
\end{equation}
Note, that in the time domain one has
\begin{equation}\label{series-t}
  \Lambda(t) = N(t) + [N \ast Q](t) + [N \ast Q \ast Q](t) + \ldots \ .
\end{equation}
Hence $\Lambda(t)$ is fully characterized in terms of two families of maps $N(t)$ and $Q(t)$. Note, that if both maps $N(t)$ and $Q(t)$ are CP, then $\Lambda(t)$ is CP as well since each term in (\ref{series-t}) is CP. But what about trace-preservation. Note, that
 $ \widetilde{K}_{\rm NZ}(s) \widetilde{N}(s) = \widetilde{B}(s) \widetilde{N}(s) - \widetilde{Z}(s) \widetilde{N}(s)$
and hence in the time domain
 $ [K_{\rm NZ} \ast N](t) = Q(t) + \dot{N}(t)$, and due to the fact that
$K_{\rm NZ}(t)$ kills the trace one has
\begin{equation}\label{TR}
 {\rm Tr}[Q(t) \rho] + \frac{d}{dt} {\rm Tr}[N(t) \rho] = 0  .
\end{equation}
Since $Q(t)$ is CP it implies ${\rm Tr}[Q(t) \rho] \geq 0$ and hence
 $\frac{d}{dt} {\rm Tr}[N(t) \rho] \leq 0$
for all density operators $\rho$. {Note that $N(t)$ provides a family of monotonic quantum operations \cite{QIT}. 
Summarising: if $N(t)$ is a monotonic quantum operation (with $N(0)=\oper$) and $Q(t)$ is CP  and satisfy (\ref{TR}), then (\ref{series-t}) provides a legitimate dynamical map being a solution of non-local memory equation (\ref{NZ}).} In what follows we  call $(N(t),Q(t))$ a {\em legitimate pair}. The crucial properties of legitimate pairs are summarised in Proposition 1-4:

\begin{Proposition}[Convexity] \label{PRO-1} If $\{N_k(t),Q_k(t)\}$ are  legitimate pairs then a convex combination $N(t) = \sum_k p_k N_k(t)$ and $Q(t) = \sum_k p_k Q_k(t)$ provide a legitimate pair.
\end{Proposition}
Proof: it is clear that both $N(t)$ and $Q(t)$ are CP and $N(0) = \oper$. Moreover one immediately verifies (\ref{TR}). Finally, observe that $||\widetilde{Q}(s)||_1 \leq \sum_k p_k  ||\widetilde{Q}_k(s)||_1 \leq 1$ which guarantee convergence of $[\oper - \widetilde{Q}(s)]^{-1}$.
Another interesting property reads

\begin{Proposition}[Reduced pair] \label{PRO-2} Suppose that $\{\mathbf{N}(t),\mathbf{Q}(t)\}$ defines a legitimate pair for the evolution in $\mathcal{H} \ot \mathcal{H}_E$. Then for arbitrary state $\omega$ in $\mathcal{H}_E$
\begin{equation*}\label{}
  N(t)[\rho] = {\rm Tr}_E (\mathbf{N}(t)[\rho \ot \omega]) ,\
  Q(t)[\rho] = {\rm Tr}_E (\mathbf{Q}(t)[\rho \ot \omega]) ,
\end{equation*}
provide a legitimate pair $\{{N}(t),{Q}(t)\}$ corresponding to the Hilbert space $\mathcal{H}$.
\end{Proposition}
Proof: by construction $N(t)$ and $Q(t)$ are CP and $N(0) = \oper$. Direct calculation easily verifies (\ref{TR}).
 This construction provides a straightforward generalization of
\begin{equation}\label{}
  \Lambda(t)[\rho] = {\rm Tr}_E ( e^{-i \mathbf{H} t} \rho \ot \omega  e^{i \mathbf{H} t} ) ,
\end{equation}
corresponding to $\mathbf{B}=0$. Note that $\Lambda(t)$ is a dynamical map whereas $N(t)$ is CP but it is not trace preserving but it satisfies $\frac{d}{dt} {\rm Tr}( N(t)[\rho]) \leq 0$ for all density operators $\rho$ in $\mathcal{H}$.

\begin{Proposition}[Gauge transformations] \label{PRO-3} If $\{N(t),Q(t)\}$ is a  legitimate pair and $\mathcal{F}(t)$ is a dynamical map, then
\begin{equation}\label{gauge}
  N'(t) = \mathcal{F}(t) N(t) \ ; \ \ Q'(t) = \mathcal{F}(t) Q(t)  ,
\end{equation}
provide a legitimate pair as well.
\end{Proposition}
Proof: obviously $N'(t)$ and $Q'(t)$ are CP and $N'(0) = \oper$. Simple calculation shows that $\{N'(t),Q'(t)\}$ verifies (\ref{TR}).
Hence a convex set of legitimate pairs is closed with respect to gauge transformations (\ref{gauge}). Finally, one has

\begin{Proposition}[CP shift] \label{PRO-4} If $\{N(t),Q(t)\}$ is a  legitimate pair and $\mathcal{G}(t)$ is a linear map such that $\int_0^t \mathcal{G}(\tau)d\tau$ is CP, then
\begin{equation}\label{gauge}
  N'(t) =  N(t) +  \int_0^t \mathcal{G}(\tau)d\tau\ ; \ \ Q'(t) = Q(t) -\mathcal{G}(t)  ,
\end{equation}
define a legitimate pair provided $Q'(t)$ is CP.
\end{Proposition}
The proof is again straightforward. Note, that one may take for example $\mathcal{G}(t) = p(t) Q(t)$ with $0 \leq p(t) \leq 1$.
Interestingly, using $\{N(t),Q(t)\}$--parametrization one may rewrite original equation (\ref{NZ}) in the following form
\begin{equation}\label{NEW}
  \dot{\rho}(t) = \int_0^t K(\tau) \rho(t-\tau) d\tau + \dot{N}(t)\rho_0 \ ,
\end{equation}
where the new kernel $K(t)$ is defined in terms of LT as follows
\begin{equation}\label{K=NQN}
  \widetilde{K}(s)  = s \widetilde{N}(s) \widetilde{Q}(s)  \widetilde{N}^{-1}(s) ,
\end{equation}
that is, $\widetilde{K}(s)$ and $s\widetilde{Q}(s)$ are related by a similarity transformation. In particular if $\widetilde{N}(s)$ and $\widetilde{Q}(s)$ commute, then
 $ \widetilde{K}(s)  = s  \widetilde{Q}(s)$,
or equivalently in the time domain
 $  {K}(t) = \dot{Q}(t) + \delta(t) Q(0)$.
It proves that the same evolution $\rho(t)$ may be described by two different non-local master equations: homogeneous Nakajima-Zwanzgig eq. (\ref{NZ}) or inhomogeneous (containing $\rho_0$) eq. (\ref{NEW}). Note, that the kernel $K(t)$ in (\ref{NEW}) has much simpler structure that $\widetilde{K}_{\rm NZ}(s) = s \oper - [1-\widetilde{Q}(s)]  \widetilde{N}^{-1}(s)$.

\vspace{.2cm}

{\em Markovian semigroup} --- Let $B[\rho] = \sum_\alpha K_\alpha\rho K_\alpha^\dagger$ be a CP map and consider a semigroup $N(t) = e^{-Zt}$ defined by
 $ Z[\rho] = i [C\rho - \rho C^\dagger]$,
where $C = H - \frac i2 \sum_\alpha K_\alpha^\dagger K_\alpha$ with Hermitian $H$. Note, that $N(t)[\rho] = e^{-i C t} \rho e^{i C^\dagger t}$ is CP.
Finally, let us introduce the following CP map $Q(t) = B N(t)$. One easily verifies (\ref{TR}): indeed ${\rm Tr}[(B- Z)N(t)[\rho]]=0$ due to the fact that $\mathcal{L} = B-Z$ kills the trace. Hence, $(N(t) = e^{-Zt},Q(t)=Be^{-Zt})$ provides a legitimate pair and the formula (\ref{!})  implies
$$  \widetilde{\Lambda}(s) = \frac{1}{s+Z} \left[ 1 - B\frac{1}{s+Z} \right]^{-1} = \frac{1}{s- \mathcal{L}} , $$
and hence $\Lambda(t) = e^{\mathcal{L}t}$ defines a Markovian semigroup with $\mathcal{L} = B-Z$ being the standard GKSL generator
$$   \mathcal{L}[\rho] = -i[H,\rho] + \sum_\alpha (K_\alpha \rho K_\alpha^\dagger - \frac 12 \{ K_\alpha^\dagger K_\alpha,\rho\}) . $$
Let us observe that formula (\ref{K=NQN}) gives $\widetilde{K}(s) = s \widetilde{N}(s) B$ or in the time domain $K(t) = \dot{N}(t) B + \delta(t) B$ and one may rewrite eq. (\ref{NEW}) as follows
\begin{equation}\label{gr}
  \dot{\rho}(t) = B \rho(t) - Z \int_0^t e^{-Z \tau} B {\rho}(t-\tau) d\tau - Z e^{-Z t}\rho_0 ,
\end{equation}
which is a highly nonstandard equation for Markovian evolution $\rho(t) = e^{\mathcal{L}t} \rho_0$. Note, that if $B=0$, then (\ref{gr}) reduces to the von Neumann equation $\dot{\rho}(t) = -Z \rho(t)$. Observe, that taking another GKSL generator $\mathcal{L}'$ and defining $N(t) = e^{- Zt + \mathcal{L}'t}$ and $Q(t)= B N(t)$ one finds
$   \Lambda(t) = \exp([\mathcal{L} + \mathcal{L}']t )$.

\vspace{.2cm}

{\em Reducing semigroup.} --- Consider a GKSL generator $\mathbf{L}= \mathbf{B} - \mathbf{Z}$ acting on $\mathcal{B}(\mathcal{H}\ot \mathcal{H}_E)$.
Due to Proposition \ref{PRO-2}  for arbitrary state $\omega$ in $\mathcal{H}_E$
\begin{equation}\label{}
  N(t)[\rho] = {\rm Tr}_E ( e^{-\mathbf{Z}t}[\rho \ot \omega]) ,
\end{equation}
and
\begin{equation}\label{}
  Q(t)[\rho] = {\rm Tr}_E (\mathbf{B}e^{-\mathbf{Z}t}[\rho \ot \omega]) ,
\end{equation}
provide a legitimate pair in $\mathcal{B}(\mathcal{H})$. Clearly, $\{N(t),Q(t)\}$ no longer defines a semigroup.

\vspace{.2cm}

{\em Semi-Markov evolution}. -- It is clear that replacing a family of CPTP maps $\Lambda(t)$ by a family of stochastic matrices $T(t)$ with $T(0)= \mathbb{I}$ we may formulate the corresponding classical problem
\begin{equation}\label{}
  \dot{T}(t) = \int_0^t W(\tau) T(t-\tau) d\tau ,
\end{equation}
with $W(t)$ being a classical memory kernel. A well studied example of such evolution is provided by so-called semi-Markov process or equivalently continuous time random walk \cite{Gil,B-V,NJP-2011}. It is uniquely determined by a  matrix $q_{ij}(t)$ giving rise to a collection of survival probabilities
 $g_j(t) = 1 - \int_0^t \sum_{i} q_{ij}(\tau) d\tau$. 
Defining a diagonal matrix $N(t)$ with diagonal elements $g_i(t)$ and the matrix $Q(t)$ with elements $q_{ij}(t)$ one immediately verifies classical analog of (\ref{TR}), that is,
\begin{equation}\label{TR-semi}
     \sum_j q_{ij}(t)   +    \dot{g}_i(t)  = 0 .
\end{equation}
In terms of a pair $\{N(t),Q(t)\}$ the corresponding solution for a stochastic map $T(t)$ reads
 $ \widetilde{T}(s) = \widetilde{N}(s)[\mathbb{I} - \widetilde{Q}(s)]^{-1}$, 
provided that the series $\mathbb{I} + \widetilde{Q}(s) + \widetilde{Q}(s) \widetilde{Q}(s) + \ldots$ converges. Note, that the convergence is guaranteed by $||\widetilde{Q}(s)||_1 < 1$ which is implied by the condition  $ \sum_j \int_0^\infty q_{ij}(\tau) d\tau =1$. Note, that the corresponding memeory kernel $W(t)$ is defined by
 $ W_{ij}(t) = B_{ij}(t) - \delta_{ij} Z_j(t)$,
with $Z_j(t) = \sum_k B_{kj}(t)$ and $B_{ij}(t)$ is defined in terms of LT as follows $\widetilde{B}_{ij}(s) = \widetilde{q}_{ij}(s)/\widetilde{g}_j(s)$. It proves that a classical semi-Markov evolution is fully determined by a legitimate pair $\{N(t),Q(t)\}$ (actually, in this case $N(t)$ is determined by $Q(t)$).
\vspace{.2cm}

{\em Quantum semi-Markov evolution}. -- The quantum analog of a classical semi-Markov evolution \cite{B-V,NJP-2011} consists of the legitimate pair $\{N(t),Q(t)\}$ such that
there exists an orthonormal basis $\{|1\>,\ldots,|d\>\}$ in $\mathcal{H}$ such that $N(t)[|k\>\<k|] = \sum_{\ell} n_{k\ell}(t)|\ell\>\<\ell|$. Hence $N(t)$ acts on the diagonal part of $\rho$ as the classical map. A simple example is provided by the Hadamard product $N(t)[\rho] = \sum_{i,j} a_{ij}(t) \rho_{ij} |i\>\<j|$ where $a_{ij}(t)$ defines a positive matrix. In this case $N(t)[|k\>\<k|] = a_{kk}(t) |k\>\<k|$. Note that defining $q_{ij}(t) = {\rm Tr}(|i\>\<i| Q(t)[|j\>\<j|)$ condition (\ref{TR}) reads $ \sum_j q_{ij}(t)   +    \dot{g}_i(t)  = 0 $, that is, it has exactly the same form as for the classical semi-Markov evolution (\ref{TR-semi}).

\begin{Example}  \label{E-1}
Consider the simplest case of quantum semi-Markov evolution defined by $N(t) = g(t) \oper$, with $g(t)$ being a survival probability, and $Q(t) = f(t) \mathcal{E}$, where $\mathcal{E}$ is an arbitrary quantum channel and $f(t)$ the  corresponding  waiting time distribution $f(t) = - \dot{g}(t) \geq 0$ \cite{Budini}. In this case one has $N(t)[|k\>\<k|] = g(t) |k\>\<k| $ for every orthonormal basis $\{|1\>,\ldots,|d\>\}$ in $\mathcal{H}$. One easily verifies condition (\ref{TR}) and finds for the memory kernel
\begin{equation}\label{K-semi}
  K_{\rm NZ}(t) = \kappa(t) (\mathcal{E} - \oper) ,
\end{equation}
where the function $\kappa(t)$ is defined in terms of LT by
 $ \widetilde{\kappa}(s) = s\widetilde{f}(s)/[1-\widetilde{f}(s)]$.

\end{Example}
This example shows that our conditions for $\{N(t),Q(t)\}$ are sufficient but not necessary. Consider a channel $\mathcal{E}$ being a projection, i.e. $\mathcal{E}^2=\mathcal{E}$. A simple example is provided by
 $ \mathcal{E}\rho = \rho_* {\rm Tr}\rho$,
where $\rho_*$ is a fixed density operator, that is, all density operators are mapped to $\rho_*$.  Then (\ref{series-t}) gives
\begin{equation}\label{Lambda-g}
  \Lambda(t) = g(t) \oper +  [1-g(t)] \mathcal{E}\ ,
\end{equation}
that is, a convex combination of $\oper$ and $\mathcal{E}$. Note, however, that $f(t)$ needs not be positive one needs only
$  0 \leq \int_0^t f(\tau) d\tau \leq 1$ for all $t \geq 0$. Take for example an oscillating function $f(t) = \frac 12 \omega \sin(\omega t)$ satisfying $  0 \leq \int_0^t f(\tau) d\tau \leq 1$. One has for the map (\ref{Lambda-g}): $\Lambda(t) = \frac 12 [1 + \cos(\omega t)] \oper +  \frac 12 [1 - \cos(\omega t)]\mathcal{E}$. The corresponding memory function $\kappa(t)$ reads $\kappa(t) = \frac 12 \omega^2\cos(\omega t/\sqrt{2})$.

\vspace{.2cm}



{\em Collision model.} --- The master equation (\ref{CM-ME}) derived via collision model  \cite{Palma} may be easily solved for the LT of the corresponding dynamical map
\begin{equation}\label{}
  \widetilde{\Lambda}(s) = \frac{\widetilde{\mathcal{F}}(s+\Gamma)}{\oper - \Gamma \widetilde{\mathcal{F}}(s+\Gamma)} .
\end{equation}
Note that it has exactly the structure of (\ref{!}) with $\widetilde{N}(s) = \widetilde{\mathcal{F}}(s+\Gamma)$ and $\widetilde{Q}(s) = \Gamma \widetilde{\mathcal{F}}(s+\Gamma)$. Going back to the time domain one finds $N(t) = e^{-\Gamma t} \mathcal{F}(t)$ and $Q(t) = \Gamma N(t)$.  One easily verifies (\ref{TR}). Note, that
$$   || \widetilde{Q}(s)||_1 = \Gamma || \widetilde{\mathcal{F}}(s+\Gamma) ||_1 = \frac{\Gamma}{s+\Gamma} < 1 $$
for $s > 0$. Hence, $\{N(t),Q(t)\}$ for the collision model \cite{Palma} provides  a legitimate pair.

The non-local master equation (\ref{CM-ME}) derived from the collision
was generalized in \cite{basano-1} as follows
\begin{eqnarray}\label{Eq-basano}
  \dot{\rho}(t) &=& \int_0^t d\tau f(\tau) \mathcal{F}(\tau) \mathcal{E} \dot{\rho}(t-\tau) + g(t) \dot{\mathcal{F}}(t) \rho_0  \nonumber \\ &+& f(t) \mathcal{F}(t)[\mathcal{E} - \oper]\rho_0\ ,
\end{eqnarray}
where $\mathcal{E}$ is a quantum channel and $f(t)$, $g(t)$ are waiting time distribution and survival probability, respectively, i.e. $g(t) = 1 - \int_0^t f(\tau)d\tau$. Integrating by parts `$f(\tau) \mathcal{F}(\tau) \mathcal{E} \dot{\rho}(t-\tau)$' one arrives at
\begin{equation}\label{CM}
   \dot{\rho}(t) = \int_0^t d\tau K(\tau){\rho}(t-\tau) + \mathcal{I}(t) \rho_0 ,
\end{equation}
where
 $ K(t) = \left( \frac{d}{dt} [ f(t) \mathcal{F}(t)] + \delta(t) f(0)   \right) \mathcal{E}$,
and the inhomogeneous term $\mathcal{I}(t)$ is defined by
 $ \mathcal{I}(t) = \frac{d}{dt} [ g(t) \mathcal{F}(t)]$.
If $\mathcal{E} = \oper$ and $g(t) = e^{-\Gamma t}$, then (\ref{Eq-basano}) reduces to (\ref{CM-ME}) and the corresponding memory kernel reads
 $ K(t) = \Gamma  \frac{d}{dt} [ e^{-\Gamma t} \mathcal{F}(t)] + \Gamma \delta(t) \oper$
due to $f(t) = \Gamma  e^{-\Gamma t} $.  It should be stressed that being inhomogeneous equation (\ref{CM}) is not of the standard Nakajima-Zwanzig form (\ref{NZ}). Equation (\ref{CM}) may be easily solved in terms of the LT of the corresponding map $\Lambda_t$ \cite{basano-1}:
\begin{equation}\label{}
  \widetilde{\Lambda}(s) = \widetilde{g \mathcal{F}}(s)\,[\oper - \widetilde{f \mathcal{F}}(s) \mathcal{E}]^{-1}  .
\end{equation}
Note, that again it has exactly the form of (\ref{!}) with
$  \widetilde{N}(s) = \widetilde{g \mathcal{F}}(s)$ and  $\widetilde{Q}(s) = \widetilde{f \mathcal{F}}(s) \mathcal{E}$,
which translates to the time domain as follows
\begin{equation}\label{**}
  N(t)= g(t) \mathcal{F}(t) \ ,\ \ Q(t) =  f(t) \mathcal{F}(t) \mathcal{E} \ .
\end{equation}
Condition (\ref{TR}) is easily verified which shows that $\{N(t), Q(t)\}$ provides a legitimate pair. The corresponding memory kernel reads
\begin{equation}\label{NZ-s}
  \widetilde{K}_{\rm NZ}(s) = \frac{ \widetilde{f \mathcal{F}}(s) \mathcal{E} - [\oper -  s \widetilde{g\mathcal{F}}(s)] }{  \widetilde{g\mathcal{F}}(s) }
\end{equation}
and in general it is not feasible to invert this formula to the time domain. On the other hand the memory kernel for the inhomogeneous equation  has simple form
  $\widetilde{K}(s) = s \widetilde{Q}(s) = \widetilde{f\mathcal{F}}(s) \mathcal{E}$.
This proves that collision model may be described by inhomogeneous memory kernel equation (\ref{CM}) with relatively simple kernel $K(t)$ or equivalently by Nakajima-Zwanzig homogeneous equation (\ref{NZ}) but the price one pays for that is highly nontrivial form of the memory kernel (\ref{NZ-s}).

It should be clear that this model provides a straightforward generalization of the quantum semi-Markov model from Example \ref{E-1}.

\vspace{.2cm}

{\em Non-commutative generalization of collision model}. -- Consider time-dependent GKSL generator
\begin{equation}\label{}
  \mathcal{L}(t)[\rho] = - i[H(t),\rho] + (\Phi(t)[\rho] - \frac 12[X(t)\rho + \rho X(t)] ) ,
\end{equation}
where $\Phi(t)$ is CP, and $X(t) = \Phi^*(t)[\mathbb{I}]$. Define
\begin{equation}\label{}
  G(t)[\rho] = V(t) \rho V^\dagger(t) ,
\end{equation}
where $V(t) = \mathcal{T}\exp(i\int_0^t C(\tau) d\tau)$, with $C(t) = H(t) - \frac i2 X(t)$.
Now, let $F(t) = \Phi(t) G(t)$. It is easy to see that $\{G(t),F(t)\}$ defines a legitimate pair: indeed $G(t)$ is obviously CP and $G(0) = \oper$. Moreover, $F(t)$ is CP being composition of two CP maps $\Phi(t)$ and $G(t)$. Finally, the condition (\ref{TR}) is trivially verified. Note that $F(t)$ and $G(t)$ may be considered as non-commutative analogs of waiting time distribution and survival probability, respectively.  Indeed, defining $g(t) = {\rm Tr}(G^*(t)[\mathbb{I}])$ and $f(t) = {\rm Tr}(F^*(t)[\mathbb{I}])$, one verifies $\dot{g}(t) = - f(t)$ (we denote by $\Phi^*$ a linear map dual to $\Phi$ \cite{DUAL}). Now, for arbitrary dynamical map $\mathcal{F}(t)$ and a channel $\mathcal{E}$ define
\begin{equation}\label{}
   N(t) = \mathcal{F}(t) G(t)\ ; \ \ Q(t) = \mathcal{F}(t)\mathcal{E} F(t) .
\end{equation}
which is a non-commutative generalization of (\ref{**}).
Due to Proposition \ref{PRO-3} $\{N(t),Q(t)\}$ provides a legitimate pair which is a direct generalization of $N(t) = g(t) \mathcal{F}(t)$ and $Q(t) = f(t) \mathcal{F}(t) \mathcal{E}$ from the collision model.

\vspace{.2cm}

{\em Conclusions.} --- In this Letter we provided a parametrization of legitimate memory kernels by legitimate pairs $\{N(t),Q(t)\}$  enjoying  a series of elegant properties (Proposition 1-4). This representation covers in a natural way Markovian semigroup,  semi-Markov evolution, collision models and their generalizations. Interestingly, using this representation one may rewrite original homogeneous Nakajima-Zwanzig memory kernel master equation (\ref{NZ}) as an inhomogeneous master equation (\ref{NEW}) containing explicitly the information about the initial state $\rho_0$. It only proves that the same evolution may be governed by different master equations. Note, that Eq. (\ref{NEW}) may be interpreted as an original Nakajima-Zwanzig equation rewritten in the {\em interaction picture} with respect to the map $N(t)$. It would be interesting to link the properties of legitimate pairs $\{N(t),Q(t)\}$ with well studied properties of the corresponding dynamical maps like for example Markovianity. Note that in these scheme the role of $Q(t)$ is to provide trace-preserving $\Lambda_t$ out of not trace-preserving but already completely positive $N_t$. It might be expected that the essential properties of the evolution might be already encoded into $N_t$. This is an open problem which deserves further analysis.

{ Finally, let us compare two representations of the dynamical map: one defined by (\ref{series-t}) and the other defined by the Dyson series
\begin{equation*}\label{}
  \Lambda(t) = \oper + \int_0^t dt_1 \mathcal{L}(t_1) +  \int_0^t dt_1 \int_0^{t_1} dt_2 \mathcal{L}(t_1) \mathcal{L}(t_2) + \ldots ,
\end{equation*}
which provides a solution for the time-local master equation (\ref{ME}).  Let us stress that these two representations are complementary since they do control complementary properties of $\Lambda(t)$ --- preservation of trace and CP. The Dyson series controls preservation of trace: identity map $\oper$ is trace-preserving and all other terms kill the trace. Hence, if one approximates $\Lambda(t)$ by  taking only finite number of terms the trace-preservation remains but CP is lost. On the contrary, each term in (\ref{series-t}) is CP hence approximating $\Lambda(t)$ by truncating (\ref{series-t})  provides the map which is obviously CP but no longer trace-preserving.  Note, however, that in this case a truncated series provides a map $\Lambda_{\rm tr}(t)$ which defines  not a quantum channel but a quantum operation, that is,  $\Lambda_{\rm tr}(t)$ is CP and ${\rm Tr}[\Lambda_{\rm tr}(t) \rho] \leq {\rm Tr}[\Lambda(t) \rho] = {\rm Tr}\rho$. This is an essential advantage of {\em non-local representation} --- starting from monotonic quantum operation $N(t)$ one provides a perturbative method  such that at each step of perturbation series  (\ref{series-t}) one has legitimate quantum operation and in the limit yields the quantum channel -- legitimate dynamical map.   }

\vspace{.2cm}

\vspace{-1cm}

\section*{Acknowledgements}

This paper was partially supported by the National Science Center project UMO-2015/17/B/ST2/02026.


\begin{thebibliography}{1}
\bibitem{DUAL} A dual map $\Phi^*$ is defined by ${\rm Tr}(A \Phi[B]) = {\rm Tr}(\Phi^*[A] B)$ for all $A, B \in \mathcal{B}(\mathcal{H})$. $\Phi$ is trace-preserving iff its dual is unital, i.e. $\Phi^*[\mathbb{I}] = \mathbb{I}$
\end{thebibliography}
\end{document}